\documentclass[12pt]{article}
\usepackage{amssymb}
\usepackage{amsfonts,amssymb,amsmath,amsthm}
\usepackage{epsfig}
\usepackage{slashed}
\usepackage{graphicx}
\usepackage{mathrsfs}
\usepackage{bbm}
\def\and{{\rm and}}

\def\a{\alpha}
\def\b{\beta}
\def\e{\eta}
\def\ep{\epsilon}
\def\d{\delta}
\def\g{\gamma}

\def\l{\lambda}
\def\s{\sigma}

\def\om{\omega}
\def\p{\partial}
\def\th{\theta}
\def\ti{\tilde}

\def\L{\Lambda}

\begin{document}
\renewcommand{\thefootnote}{\fnsymbol{footnote}}
\begin{titlepage}

\vspace{10mm}
\begin{center}
{\Large\bf $U(2,2)$ gravity on noncommutative space with symplectic structure}
\vspace{16mm}

{\large Yan-Gang Miao\footnote{\em E-mail: miaoyg@nankai.edu.cn}, Zhao Xue\footnote{\em E-mail: illidanpimbury@mail.nankai.edu.cn} and
Shao-Jun Zhang\footnote{\em E-mail: sjzhang@mail.nankai.edu.cn}

\vspace{6mm}
{\normalsize \em Department of Physics, Nankai University, Tianjin 300071, \\
People's Republic of China}}

\end{center}

\vspace{10mm}
\centerline{{\bf{Abstract}}}
\vspace{6mm}

The classical Einstein's gravity can be reformulated from the
constrained $U(2,2)$ gauge theory on the ordinary (commutative) four-dimensional spacetime.
Here we consider a noncommutative manifold with a symplectic structure
and construct a $U(2,2)$ gauge theory on such a manifold by using
the covariant coordinate method.
Then we use the
Seiberg-Witten map to express noncommutative quantities in terms of their
commutative counterparts up to the first-order in noncommutative parameters. After imposing constraints we obtain a noncommutative gravity theory
described by the Lagrangian with up to nonvanishing first order corrections in noncommutative parameters.
This result coincides with our previous one obtained for the noncommutative $SL(2,C)$ gravity.

\vskip 20pt
PACS Number(s): 04.90.+e; 02.40.Gh; 11.10.Nx


\end{titlepage}

\newpage
\renewcommand{\thefootnote}{\arabic{footnote}}
\setcounter{footnote}{0}
\pagenumbering{arabic}

\section{Introduction}

The concept of noncommutative spacetimes was first introduced by
Snyder~\cite{Snyder} in order to solve the divergence
problem of the quantum field theory. However, this difficulty was not overcome completely within the framework of noncommutative spacetimes.
Although it could provide an ultraviolet cutoff, the noncommutative spacetime
gave rise to some other troubles, such as the well-known UV/IR
mixing~\cite{Minwalla}. Since the technique of renormalization was proposed,
the noncommutative attempt was not popular for a long time,
until the 1990s, when Seiberg and Witten~\cite{SW} suggested that the D-brane dynamics
under a B-field background can be described by a noncommutative
field theory, which connects the noncommutative spacetime with the string theory and also revives the idea of spacetime noncommutativity.
As it was known,
the noncommutative field theory was considered  on the other hand to be a good candidate to describe the physics with the scale less than the Planck's.
For the recent progress on the noncommutative issue, see, for instance, some reviews~\cite{Douglas,Szabo,Banerjee}.

The noncommutative formulation of gravity has been
considered~\cite{NCgravrev} to be a necessity for quantization of
gravity, and some interesting approaches have been suggested to give a noncommutative
gravity in which the noncommutative properties are presented by the Moyal-Weyl
product~\cite{Moyal}. For an overview, let us give a brief summary on the approaches.
Within the framework of the gauge theory of gravity, a kind of deformation of the gravity theory is
constructed~\cite{Chamseddine:2000} by gauging the noncommutative $SO(1,4)$ de Sitter group
and contracting it to $ISO(1,3)$ by the Seiberg-Witten
map~\cite{SW}, and another effort~\cite{Marculescu:2008}
is made to build the $SO(1,3)$ noncommutative formulation of gravity. In refs.~\cite{Chamseddine:2002,Cardellaetal:2002}
the noncommutative formulation of gravity
is realized by breaking the gauge group $U(2,2)$ into $U(1,1)\otimes U(1,1)$ in terms of constraints.
From a quite different point of view, the theory of gravity and its noncommutative extension can be expressed in a $GL(2,C)$ formulation
with complex vierbeins~\cite{Chamseddine:2003}. In ref.~\cite{Calmet} a noncommutative formulation of gravity is given
by a class of restricted diffeomorphism symmetries that preserves the noncommutative algebra.
Moreover, a gravity theory on noncommutative spaces is proposed~\cite{Wess 2005,Wess 2006} in terms of a twisted diffeomorphism algebra
from a purely geometrical point of view. We note that the noncommutative formulations of gravity mentioned above
are worked out on the so-called canonical noncommutative spacetime with constant noncommutative parameters $\th^{\mu\nu}$.

The noncommutative gravity related to coordinate-dependent noncommutative parameters
has also been discussed. For instance,
the noncommutative theory of gravity is constructed~\cite{Banerjee:2007} based on the work of ref.~\cite{Calmet} on a noncommutative spacetime
with the Lie algebraic structure, which is in fact a special case of a Poisson manifold.
Furthermore,
an $SL(2,C)$ formulation of gravity on the noncommutative space with symplectic manifolds
is proposed~\cite{miao} by us in light of a different
starting point from that of refs.~\cite{Chamseddine:2002,Cardellaetal:2002}.
Because the field strength defined by the way of refs.~\cite{Chamseddine:2002,Cardellaetal:2002} is of no
gauge invariance on the symplectic manifold, we thus utilize the covariant coordinate technique~\cite{Wessetal:20001} to construct
covariant actions in our previous work~\cite{miao} and in this work as well.

The present paper focuses on the $U(2,2)$ gravity on the noncommutative space with the symplectic structure.
In the next section, we give a brief introduction on how to
construct a gravity theory from a $U(2,2)$ gauge theory with constraints on the commutative (ordinary) spacetime.
Section 3 is the main context of this paper
and it contains three subsections.
In the first subsection, we give the gauge invariant
action of the noncommutative $U(2,2)$ gauge theory on the symplectic manifold. In the second we expand the star product to the first order
in noncommutative parameters and
calculate the lagrangian of the noncommutative $U(2,2)$ gauge theory (see eq.~(\ref{NCaction2})) which is expressed
totally by noncommutative gauge fields to the same order.
At last in the third subsection we first apply the Seiberg-Witten map\footnote{We have two intentions to apply the Seiberg-Witten map:
One is to get a noncommutative action represented completely by commutative quantities,
and the other is to use the constraints (eq.~(\ref{constraint})) for breaking the symmetry group and deleting the redundant degrees of freedom.
The latter is just a technique in calculation we adopt in the present paper. The Seiberg-Witten map ensures that we can add constraints to
the corresponding commutative quantities. Incidentally,
the Seiberg-Witten map was utilized in refs.~\cite{Chamseddine:2002,Cardellaetal:2002} in a different way from ours, i.e.
after the addition of constraints.}
to establish a relation between
noncommutative quantities and their commutative counterparts, and then by imposing constraints and breaking
the group we obtain
the noncommutative $U(2,2)$ gravity in which the Lagrangian is given up to first order corrections in noncommutative parameters.
Finally we make a conclusion in section 4.

\section{Gravity based on $U(2,2)$ gauge group}
Let us at first give a brief introduction to the gravity model based on the gauge group $U(2,2)$
with constraints~\cite{Chamseddine:2002,Cardellaetal:2002}.
The $U(2,2)$ group is a Lie group of complex $4 \times 4$ matrices $U$  satisfying the following condition:
\begin{equation}
U^\dagger \tilde{\e} U = \tilde{\e},
\end{equation}
where $\tilde{\e} = {\rm diag} (+ + - -)$. Therefore,
some basis of the corresponding Lie algebra $u(2,2)$ is given by 16 linear independent matrices $\l$
satisfying the relation:
\begin{equation}
\l^\dagger = \tilde{\e} \l \tilde{\e}.
\end{equation}
In the Dirac-Pauli representation with $\g_0 = - i \tilde{\e}$, one
can choose the following set of matrices $\tau^I$, $I=1,\dots,16$,
as the basis of $u(2,2)$:
\begin{equation}
\left(1,~ i\g_5,~ i\g_a,~ i\g_a\g_5,~ i \g_{ab}\right), \label{generators}
\end{equation}
where $\g_5 \equiv  i \g_0 \g_1 \g_2 \g_3$ and $\g_{ab}  \equiv \frac{1}{2} [\g_a, \g_b]$ with gamma matrices $\g_a$, $a=0, 1, 2, 3$,
satisfying the Clifford algebra
\begin{equation}
\left\{\g_a, \g_b\right\} = 2 \e_{ab}, \qquad \e_{ab} = {\rm diag}(- + + +),
\end{equation}
where $\{\cdot, \cdot\}$ stands for an anticommutator.

The gauge field $A_\mu$ is Lie algebra valued:
\begin{equation}
A_\mu = a_\mu + i b_\mu \g_5 + i e^a_\mu \g_a + i f^a_\mu \g_a \g_5 + \frac{i}{4} \om_\mu^{ab} \g_{ab}. \label{Adecomp}
\end{equation}
The field strength is defined to be
\begin{equation}
F_{\mu\nu} = \p_\mu A_\nu - \p_\nu A_\mu - i \left[A_\mu, A_\nu\right].
\end{equation}
It can be decomposed in terms of the $u(2,2)$ algebra generators (eq.~(\ref{generators})) as
\begin{equation}
F_{\mu\nu} = F^1_{\mu\nu} + i F^5_{\mu\nu} \g_5 + i F^a_{\mu\nu} \g_a + i F^{a5}_{\mu\nu} \g_a \g_5 + \frac{i}{4} F^{ab}_{\mu\nu} \g_{ab},
\end{equation}
where the components are given by
\begin{eqnarray}
F^1_{\mu\nu} &=& \p_\mu a_\nu - \p_\nu a_\mu,\nonumber\\
F^5_{\mu\nu} &=& \p_\mu b_\nu - \p_\nu b_\mu + 2 e^a_\mu f_{\nu a} - 2 e^a_\nu f_{\mu a},\nonumber\\
F^a_{\mu\nu} &=& \p_\mu e^a_\nu - \p_\nu e^a_\mu + \om^{ab}_\mu e_{\nu b} - \om^{ab}_\nu e_{\mu b}
+ 2 f^a_\mu b_\nu - 2 f^a_\nu b_\mu, \nonumber\\
F^{a5}_{\mu\nu} &=& \p_\mu f^a_\nu - \p_\nu f^a_\mu + \om^{ab}_\mu f_{\nu b} - \om^{ab}_\nu f_{\mu b} + 2 e^a_\mu b_\nu - 2 e^a_\nu b_\mu,\nonumber\\
F^{ab}_{\mu\nu} &=& \p_\mu \om^{ab}_\nu - \p_\nu \om^{ab}_\mu + \om^{ac}_\mu \om_{\nu c}^{~~b} - \om^{ac}_\nu \om_{\mu c}^{~~b}
+ 8 e^a_\mu e^b_\nu - 8 f^a_\mu f^b_\nu.\label{components}
\end{eqnarray}

Under an infinitesimal gauge transformation, the gauge field $A_\mu$ and its strength $F_{\mu\nu}$ transform as follows:
\begin{eqnarray}
\d_\L A_\mu &=& \p_\mu \L + i \left[\L, A_\mu\right],\\
\d_\L F_{\mu\nu} &=& i \left[\L, F_{\mu\nu}\right],
\end{eqnarray}
where $\L \equiv \L_I \tau^I$ is an infinitesimal transformation parameter. Thus it is not difficult to write a gauge invariant action
\begin{equation}
S = i \int d^4x ~\ep^{\mu\nu\rho\s} \mathrm{Tr} \left(\g_5 F_{\mu\nu} F_{\rho\s}\right). \label{action}
\end{equation}
In terms of the component expressions of $F_{\mu\nu}$ (see eq.~(\ref{components})),
together with the trace identities of the gamma matrices, the action eq.~(\ref{action})
can be rewritten as
\begin{equation}
S = - \int d^4x ~ \ep^{\mu\nu\rho\s} \left(8 F^1_{\mu\nu} F^5_{\rho\s} + \frac{1}{4} \ep_{abcd} F^{ab}_{\mu\nu} F^{cd}_{\rho\s}\right).\label{action2}
\end{equation}
When one imposes the constraints
\begin{equation}
a_\mu = b_\mu = 0, ~~~~f^a_\mu = \a e^a_\mu,~~~~F^a_{\mu\nu} = 0, \label{Constraints}
\end{equation}
which break the gauge group $U(2,2)$ into $SO(1,3)$ with an additional $U(1)$ global symmetry, the action eq.~(\ref{action2}) becomes
\begin{equation}
S= - \frac{1}{4} \int d^4x ~\ep^{\mu\nu\rho\s} \ep_{abcd} \bigg(R^{ab}_{\mu\nu} + 8 \left(1- \a^2\right) e^a_\mu e^b_\nu\bigg) \bigg(R^{cd}_{\rho\s}
+ 8 \left(1- \a^2\right) e^c_\rho e^d_\s\bigg), \label{action3}
\end{equation}
where the curvature tensor
$R^{ab}_{\mu\nu} \equiv \p_\mu \om^{ab}_\nu - \p_\nu \om^{ab}_\mu + \om^{ac}_\mu \om_{\nu c}^{~~b} - \om^{ac}_\nu \om_{\mu c}^{~~b}$.
For the case $\a = 1$, eq.~(\ref{action3}) gives the topological Gauss-Bonnet term. For the case $\a \neq 1$, it gives,
besides the topological Gauss-Bonnet term, the classical Einstein action plus a cosmological term.

In the next section, we generalize this formulation of gravity to a noncommutative space with a symplectic structure.

\section{Noncommutative version of gravity on symplectic manifold}
Consider a manifold $M$ on which a Poisson bracket is defined:
\begin{equation}
\{f(x), g(x)\}_{\rm Poisson} = \th^{\mu\nu} (x) \p_\mu f(x) \p_\nu g(x),\label{poisson}
\end{equation}
where $\th^{\mu\nu} = - \th^{\nu\mu}$ is a Poisson bivector and $f(x)$ and $g(x)$ are arbitrary functions on $M$.
The Jacobi identity of the Poisson bracket imposes the following condition on the bivector $\th^{\mu\nu} (x)$:
\begin{equation}
\th^{\mu\rho}(x) \p_\rho \th^{\nu\s}(x) + \th^{\nu\rho}(x) \p_\rho
\th^{\s\mu}(x) + \th^{\s\rho}(x) \p_\rho \th^{\mu\nu}(x) = 0. \label{Jacobi}
\end{equation}
A manifold with such a Poisson structure is called a Poisson manifold. Consider a special case in which the
functions $f(x)$ and $g(x)$ are coordinates, and we get the following relations:
\begin{equation}
\{x^\mu, x^\nu\}_{\rm Poisson} = \th^{\mu\nu} (x).
\end{equation}
In the quantum theory, the Poisson bracket is replaced by a commutator.
Then we arrive at a noncommutative manifold with the following commutation relations\footnote{Note that $\theta^{\mu\nu}$ satisfies the Jacobi identity
even though it is a function of operators. Alternatively,
when we go from the operator product to the star product along the Weyl deformation quantization procedure, higher order terms will appear:
\begin{equation*}
[x^\mu, x^\nu]_\star = W^{-1} (i \theta^{\mu\nu}(\hat{x})) = i \theta^{\mu\nu} (x) + \mathcal{O} (\theta^{\mu\nu}(x)),
\end{equation*}
where $W^{-1}$ is the inverse of the Weyl map; see ref.~\cite{Calmet:2003}.
Therefore, it is the term $i \theta^{\mu\nu} (x) + \mathcal{O} (\theta^{\mu\nu}(x))$
rather than the term $i \theta^{\mu\nu} (x)$ that satisfies the Jacobi identity. However, we do not consider the higher order terms $\mathcal{O}
(\theta^{\mu\nu}(x))$ in the present paper.
Note also that $\theta_{\mu\nu}$ has the order of $-1$; see eq.~(\ref{Borelation}) and the explanation below it. As a result,
there are no higher order terms in eqs.~(\ref{NCCF1})-(\ref{NCCFab}),
and consequently  the higher order correcting terms of $\theta^{\mu\nu}$ do not affect our results.}:
\begin{equation}
[\hat{x}^\mu, \hat{x}^\nu] = i \th^{\mu\nu} (\hat{x}).
\label{noncommutativity}
\end{equation}
As dealt with to the $SL(2,C)$ gravity in ref.~\cite{miao},
here we still suppose that the bivector $\th^{\mu\nu} (x)$ is nondegenerate; therefore, we can define its inverse $\th_{\mu\nu} (x)$ as
$\th^{\mu\nu} \th_{\nu\rho} = \d^\mu_\rho$. By using the Jacobi identity (eq.~(\ref{Jacobi})), we can show that the two-form
$\Theta = \frac{1}{2} \th_{\mu\nu} dx^\mu \wedge dx^\nu$ is closed ($d \Theta=0$) and thus prove that the manifold is symplectic.
In this paper we shall restrict our discussions on the noncommutative spacetime with the symplectic structure.

According to Kontsevich's deformation~\cite{Kontsevich}, there
exists an associative star product between functions to a given Poisson bivector
$\th^{\mu\nu} (x)$ and the star product can be written as
\begin{equation}
f(x) \star g(x) = f(x) g(x) + \frac{i}{2} \th^{\mu\nu}(x) \p_\mu
f(x) \p_\nu g(x) + \mathcal{O} (\th^2).\label{star}
\end{equation}
Note that it is not unique for higher order terms. In order to avoid this ambiguity we shall restrict our discussions
only to the first-order in $\th^{\mu\nu}(x)$. We shall see in subsection 3.3 that this restriction is consistent with the first order Seiberg-Witten map
to the noncommutative $U(2,2)$ gravity.

In the following subsections we
construct a gravity model based on the constrained $U(2,2)$ gauge group on the noncommutative spacetime depicted by
eq.~(\ref{noncommutativity}) with the symplectic structure mentioned above.

\subsection{Construction of noncommutative gravity}
Because of coordinate dependence of $\th^{\mu\nu}(x)$, we cannot define a gauge field strength which transforms
covariantly simply by using the method adopted in the commutative case. Here we can follow the covariant coordinate approach\footnote{Because of the
coordinate dependence of the Poisson tensor, it is hard to define the covariant derivative straightforwardly. Despite
some work following the covariant derivative approach~\cite{Calmet:2003}, the covariant coordinate approach that has been
utilized in refs.~\cite{Wessetal:20001,Banerjee:2007} is quite straightforward. Here we find that it is an easy way to realize our idea.}
which was proposed in ref.~\cite{Wessetal:20001} and has been applied~\cite{miao} by us to the $SL(2,C)$ gravity. The covariant coordinate is defined as
\begin{equation}
\hat{X}^\mu = x^\mu\textbf{1} + \hat{B}^\mu, \label{Covcoord}
\end{equation}
where all the quantities are matrices, and it complies with the gauge transformation:
\begin{equation}
\d_{\hat{\L}} (\hat{X}^\mu \star \hat{\Psi}) = i \hat{\L} \star (\hat{X}^\mu \star \hat{\Psi}),\label{NCcovcortran}
\end{equation}
where $\hat{\Psi}$ is an arbitrary matter field with the gauge transformation
\begin{equation}
\d_{\hat{\L}} \hat{\Psi} = i \hat{\L} \star \hat{\Psi} \label{NCmatttran}.
\end{equation}
From eqs.~(\ref{NCcovcortran}) and (\ref{NCmatttran}), we get the gauge transformations of the field $\hat{B}^\mu$:
\begin{eqnarray}
\d_{\hat{\L}} \hat{B}^\mu &=& i [\hat{\L}, x^\mu]_\star + i [\hat{\L}, \hat{B}^\mu]_\star \nonumber\\
&=& \th^{\mu\nu} \p_\nu \hat{\L} + i [\hat{\L}, \hat{B}^\mu]_\star,\label{NCBtran}
\end{eqnarray}
and that of the covariant coordinate $\hat{X}^\mu$ by using eq.~(\ref{Covcoord}),
\begin{equation}
\d_{\hat{\L}} \hat{X}^\mu = i [\hat{\L}, {\hat{X}}^{\mu}]_\star. \label{CovCoorTran}
\end{equation}
The noncommutative gauge filed $\hat{A}_\mu$ is defined as~\cite{Banerjee:2007,Wessetal:20001}
\begin{equation}
\hat{A}_\mu = \th_{\mu\nu} \hat{B}^\nu, \label{Borelation}
\end{equation}
where $\th_{\mu\nu}$ is the inverse of $\th^{\mu\nu}$ and can be considered as of order $({\th^{\mu\nu}})^{-1}$
when we count the power of $\th^{\mu\nu}$.
Using eq.~(\ref{NCBtran}) and eq.~(\ref{Borelation}),
we can derive the gauge transformation of the gauge field $\hat{A}_\mu$ up to the first order in $\th^{\mu\nu}$:
\begin{equation}
\d_{\hat{\L}} \hat{A}_\mu = \p_\mu \hat{\L} + i [\hat{\L}, \hat{A}_\mu] - \frac{1}{2} \th^{\l\s} \{\p_\l \hat{\L}, \p_\s \hat{A}_\mu\}
- \frac{1}{2} \th_{\mu\a} \th^{\l\s} \p_\s \th^{\a\b} \{\p_\l \hat{\L}, \hat{A}_\b\}. \label{NCAtran}
\end{equation}

In light of the covariant coordinate approach~\cite{Wessetal:20001},
we first define a rank-two tensor $\hat{F}^{\mu\nu}$ composed of the covariant coordinates
and of the noncommutative parameters in order to obtain the field strength,
\begin{equation}
\hat{F}^{\mu\nu} = - i ([\hat{X}^\mu, \hat{X}^\nu]_\star - i \th^{\mu\nu} (\hat{X})), \label{SecTensor}
\end{equation}
where $\th^{\mu\nu} (\hat{X})$ is the Poisson tensor (eq.~(\ref{poisson})) in which $x$ has been replaced by $\hat{X}$
in order for $\hat{F}^{\mu\nu}$ to be
a function of covariant coordinates. Because of eq.~(\ref{CovCoorTran}) and the gauge transformation
of $\th^{\mu\nu} (\hat{X})$, i.e. $\d_{\hat{\L}} \th^{\mu\nu} (\hat{X}) = i [\hat{\L}, \th^{\mu\nu} (\hat{X})]_\star$,
the gauge transformation of this rank-two tensor $\hat{F}^{\mu\nu}$ takes the form
\begin{equation}
\d_{\hat{\L}} \hat{F}^{\mu\nu} = i [\hat{\L}, \hat{F}^{\mu\nu}]_\star. \label{covar}
\end{equation}
Now it is time to look for the relation between the rank-two tensor $\hat{F}^{\mu\nu}$ and the gauge field strength $\hat{F}_{\mu\nu}$.
In the case of the canonical noncommutative space where the noncommutative parameters are constant, the relation is trivial:
$\hat{F}^{\mu\nu} = \th^{\mu\rho} \th^{\nu\s} \hat{F}_{\rho\s}$. But in our case where the parameters are coordinate-dependent,
we should modify the relation
to ensure that the gauge field strength $\hat{F}_{\mu\nu}$ transforms covariantly:
\begin{equation}
\d_{\hat{\L}} \hat{F}_{\mu\nu} = i [\hat{\L}, \hat{F}_{\mu\nu}]_\star \label{NCRtran}.
\end{equation}
In order to achieve this goal, one can introduce~\cite{Calmet:2003} such a function $\hat{\th}_{\mu\nu} (\hat{X})$ that has the following transformation:
\begin{equation}
\d_{\hat{\L}} \hat{\th}_{\mu\nu} (\hat{X}) = i [\hat{\L}, \hat{\th}_{\mu\nu}]_\star.\label{deltatheta}
\end{equation}
Consequently, if it is defined by\footnote{Because of the noncommutativity of the star product there are other possibilities to define the field
strength, such as
$\hat{F}_{\mu\nu} \equiv \hat{\th}_{\mu\l} \star \hat{F}^{\l\s} \star \hat{\th}_{\nu\s}$ or
$\hat{F}_{\mu\nu} \equiv \hat{F}^{\l\s} \star \hat{\th}_{\mu\l} \star \hat{\th}_{\nu\s}$. Here we just choose the most understandable form.}
\begin{equation}
\hat{F}_{\mu\nu} \equiv \hat{\th}_{\mu\l} \star \hat{\th}_{\nu\s} \star \hat{F}^{\l\s}, \label{NCRrelation}
\end{equation}
the gauge field strength satisfies the transformation property eq.~(\ref{NCRtran}).
In the next subsection, we can see the function $\hat{\th}_{\mu\nu} (\hat{X})$
indeed exists and we shall give its expansion expression.

Now it is straightforward for us to write a gauge invariant action on the noncommutative spacetime with the symplectic structure:
\begin{equation}
S = i \int d^4 x ~\left(\mathrm{det} \th^{\mu\nu}\right)^{- \frac{1}{2}} \ep^{\mu \nu \rho \s}
{\rm Tr} \left({\g _5} \hat{F}_{\mu\nu} \star \hat{F}_{\rho\s}\right), \label{NCaction}
\end{equation}
where the symplectic volume form $(\mathrm{det} \th^{\mu\nu})^{- \frac{1}{2}} d^4x$
appears naturally.\footnote{$(\det{\theta^{\mu\nu}})^{-\frac{1}{2}}d^4x$ is the natural volume
form of a symplectic manifold, like the Liouville measure ${(2\pi \hbar)}^{-1}{dp dq}$ of the phase space.
Therefore the geometric meaning is obvious.}
This factor guarantees the trace property of the integral~\cite{Felder:2000,Calmet:2003,miao}:
\begin{equation}
\int d^4 x ~\left(\mathrm{det} \th^{\mu\nu}\right)^{- \frac{1}{2}} f(x) \star g(x) = \int d^4 x ~\left(\mathrm{det} \th^{\mu\nu}\right)^{- \frac{1}{2}}
g(x) \star f(x) = \int d^4 x ~\left(\mathrm{det} \th^{\mu\nu}\right)^{- \frac{1}{2}} f(x) g(x),
\end{equation}
where $f(x)$ and $g(x)$ are arbitrary functions. With this property, it is easy to prove the gauge invariance of the action (eq.~(\ref{NCaction})).

\subsection{First-order approximation}
On the basis of the action of noncommutative gravity on the noncommutative manifold,
we now compute the first-order correction for the Lagrangian by using the expansion
of the star product and the Seiberg-Witten map.
From eq.~(\ref{NCaction}) we know that it is enough to calculate the gauge field strength $\hat{F}_{\mu\nu}$
up to the first order in $\th^{\mu\nu}$ (see eqs.~(\ref{NCCF1})-(\ref{NCCFab}) or their Seiberg-Witten maps upon which constraints imposed,
eqs.~(\ref{NCCF12})-(\ref{NCCFab2})).

First let us express $\hat{F}^{\mu\nu}$ in terms of $\hat{A}_\mu(x)$ and $\th^{\mu\nu}(x)$ by substituting eqs.~(\ref{Covcoord}) and (\ref{Borelation})
into eq.~(\ref{SecTensor}) and expanding $\th^{\mu\nu} (\hat{X})$ to the third order in $\th^{\mu\nu}(x)$\footnote{The purpose of expanding
$\th^{\mu\nu} (\hat{X})$ to the third order in $\th^{\mu\nu}(x)$ is to ensure the following expansion of the gauge field strength $\hat{F}_{\mu\nu}$
up to the first order in $\th^{\mu\nu}$. See eq.~(\ref{NCRrelation}).}:
\begin{eqnarray}
\hat{F}^{\mu\nu} &= &\th^{\mu\l} \th^{\nu\s} (\p_\l \hat{A}_\s - \p_\s \hat{A}_\l - i [\hat{A}_\l, \hat{A}_\s])
+ \frac{1}{2} \th^{\mu\l} \th^{\nu\s} \th^{\d \e} \{\p_\d \hat{A}_\l, \p_\e \hat{A}_\s\} \nonumber\\
&& + \frac{1}{2} \th^{\mu\l} \p_\e \th^{\nu\s} \th^{\d\e} \{\p_\d \hat{A}_\l, \hat{A}_\s\}
+ \frac{1}{2} \p_\d \th^{\mu\l} \th^{\nu\s}\th^{\d\e} \{\hat{A}_\l, \p_\e \hat{A}_\s\}\nonumber\\
&& + \frac{1}{2} \p_\d \th^{\mu\l} \p_\e \th^{\nu\s} \th^{\d \e} \{\hat{A}_\l, \hat{A}_\s\}
+ \frac{1}{4} \th^{\l\a} \th^{\s\b} \p_\l \p_\s \th^{\mu\nu} \{\hat{A}_\a, \hat{A}_\b\}, \label{NCF1}
\end{eqnarray}
where $\hat{A}_\mu$ can be expressed in terms of the $u(2,2)$ algebra generators to be
\begin{equation}
\hat{A}_\mu = \hat{A}^1_\mu + i \hat{A}^5_\mu \g_5 + i \hat{A}^a_\mu
\g_a + i \hat{A}^{a5}_\mu \g_a \g_5 + \frac{i}{4} \hat{A}_\mu^{ab}
\g_{ab}\label{Amu}.
\end{equation}
In order to have the expansion of the field strength, according to its definition, eq.~(\ref{NCRrelation}), we need
the expansion of the tensor $\hat{\theta}_{\mu\nu}$. The latter
has been given in ref.~\cite{Calmet:2003} as follows\footnote{In order to derive the first order in $\theta^{\mu\nu}$ for the field strength,
it is enough to expand $\hat{\theta}_{\mu\nu}$ to the zeroth order
in $\theta^{\mu\nu}$. This is a quite usual treatment, see  ref.~\cite{Calmet:2003} for details.}:
\begin{equation}
\hat{\th}_{\mu\nu} = \th_{\mu\nu} + \th^{\rho\s} \p_\rho
\th_{\mu\nu} \hat{A}_\s + \mathcal{O} (\th^{\mu\nu})
\label{NCtheta}.
\end{equation}
Note that in order to make the Lagrangian of the noncommutative $U(2,2)$
gauge theory (see eq.~(\ref{NCaction2})) be written totally by noncommutative quantities we have replaced $A_\s$ with $\hat{A}_\s$
in the above equation if comparing with the original equation given in ref.~\cite{Calmet:2003}.
This is only a technique in calculation which will not affect our final result
due to the Seiberg-Witten map between $\hat{A}_\s$ and $A_\s$ (see eq.~(\ref{OmMap})),
and due to the corrections for the gauge field strength $\hat{F}_{\mu\nu}$ just up to the first order in $\th^{\mu\nu}$ as well.

Second, with eqs.~(\ref{NCRrelation}), (\ref{NCF1}) and (\ref{NCtheta}) we can calculate the noncommutative field strength $\hat{F}_{\mu\nu}$
in terms of noncommutative gauge fields up to the first order in $\th^{\mu\nu}$ as follows:
\begin{eqnarray}
\hat{F}_{\mu\nu} &=& \ti{F}_{\mu\nu} + \frac{1}{2} \th^{\a\b} \{\p_\a \hat{A}_\mu, \p_\b \hat{A}_\nu\}
- \frac{1}{2} \th_{\nu\s} \p_\b \th^{\l\s} \th^{\a\b} \{\p_\a \hat{A}_\mu, \hat{A}_\l\}
+ \frac{1}{2} \th_{\mu\l} \p_\a \th^{\l\s} \th^{\a\b} \{\hat{A}_\s, \p_\b \hat{A}_\nu\} \nonumber\\
&& + \frac{1}{2} \th_{\mu\rho} \th_{\nu\s} \p_\d \th^{\rho\a} \p_\e \th^{\s\b} \th^{\d\e} \{\hat{A}_\a, \hat{A}_\b\}
+ \frac{1}{4} \th_{\mu\rho} \th_{\nu\s} \th^{\d\a} \th^{\e\b} \p_\d \p_\e \th^{\rho\s} \{\hat{A}_\a, \hat{A}_\b\} \nonumber\\
&&+ \frac{i}{2} \th^{\l\d} \p_\l \th_{\mu\rho} \p_\d \th^{\rho\a} \ti{F}_{\a\nu}
+ \frac{i}{2} \th^{\l\d} \p_\l \th_{\nu\s} \p_\d \th^{\s\b} \ti{F}_{\mu\b}
+ \frac{i}{2} \th^{\l\d} \p_\l \th_{\mu\rho} \th^{\rho\a} \p_\d \ti{F}_{\a\nu} \nonumber\\
&&+ \frac{i}{2} \th^{\l\d} \p_\l \th_{\nu\s} \th^{\s\b} \p_\d \ti{F}_{\mu\b}
+ \frac{i}{2} \th^{\l\d} \p_\l \th_{\mu\rho} \p_\d \th_{\nu\s}
\th^{\rho\a} \th^{\s\b} \ti{F}_{\a\b} + \th^{\l\d} \p_\l
\th_{\mu\rho} \th^{\rho\a} \hat{A}_\d \ti{F}_{\a\nu} \nonumber\\
&&+ \th^{\l\d}
\p_\l \th_{\nu\s} \th^{\s\b} \hat{A}_\d \ti{F}_{\mu\b}, \label{NCF2}
\end{eqnarray}
where $\ti{F}_{\mu\nu} \equiv \p_\mu \hat{A}_\nu - \p_\nu \hat{A}_\mu - i [\hat{A}_\mu, \hat{A}_\nu]$ and it can be decomposed in
terms of the $u(2,2)$ algebra generators:
\begin{equation}
\ti{F}_{\mu\nu} = \ti{F}^1_{\mu\nu} + i \ti{F}^5_{\mu\nu} \g_5 +
i \ti{F}^a_{\mu\nu} \g_a + i \ti{F}^{a5}_{\mu\nu} \g_a \g_5 +
\frac{i}{4} \ti{F}^{ab}_{\mu\nu} \g_{ab}\label{Ftilde}.
\end{equation}

Since we consider a $U(2,2)$ gauge theory on noncommutative spaces,
we can decompose $\hat{F}_{\mu\nu}$ into the following form in terms of $u(2,2)$ algebra generators:
\begin{equation}
\hat{F}_{\mu\nu} = \hat{F}^1_{\mu\nu} + i \hat{F}^5_{\mu\nu} \g_5 +
i \hat{F}^a_{\mu\nu} \g_a + i \hat{F}^{a5}_{\mu\nu} \g_a \g_5 +
\frac{i}{4} \hat{F}^{ab}_{\mu\nu} \g_{ab}.
\end{equation}
Here we introduce some notions to represent long formulas for the sake of convenience. Suppose $G$ and
$H$ are two quantities valued in the algebra $u(2,2)$:
\begin{eqnarray}
G &=& G^1 + i G^5 \g_5 + i G^a \g_a + i G^{a5} \g_a \g_5 + \frac{i}{4} G^{ab} \g_{ab},\\
H &=& G^1 + i H^5 \g_5 + i H^a \g_a + i H^{a5} \g_a \g_5 + \frac{i}{4} H^{ab} \g_{ab}.
\end{eqnarray}
When we define a function
\begin{equation}
P (G,H) \equiv G H = P^1 + i P^5 \g_5 + i P^a \g_a + i P^{a5} \g_a
\g_5 + \frac{i}{4} P^{ab} \g_{ab},
\end{equation}
we can use the Clifford algebra to compute all the components of $P(G,H)$:
\begin{eqnarray}
P^1 (G,H) &=& G^1 H^1 - G^5 H^5 - G^a H_a + G^{a5} H^5_a + \frac{1}{8} G^{ab} H_{ab},\\
P^5 (G,H) &=& G^1 H^5 + G^5 H^1 + i G^a H^5_a - i G^{a5} H_a + \frac{1}{16} G^{ab} H^{cd} \ep_{abcd},\\
P^a (G,H) &=& G^1 H^a + G^a H^1 - i G^5 H^{a5} + i G^{a5} H^5 - \frac{i}{2} G_b H^{ab} + \frac{i}{2} G^{ab} H_b \\
&&- \frac{1}{4} G^{d5} H^{bc} \ep_{bcd}^{~~~a} - \frac{1}{4} G^{bc} H^{d5} \ep_{bcd}^{~~~a},\\
P^{a5} (G,H) &=& G^1 H^{a5} + G^{a5} H^1 - i G^5 H^a + i G^a H^5 + \frac{i}{2} G^{ab} H^5_b - \frac{i}{2} G^5_b H^{ab}\\
&& - \frac{1}{4} G^d H^{bc} \ep_{bcd}^{~~~a} - \frac{1}{4} G^{bc} H^d \ep_{bcd}^{~~~a},\\
P^{ab} (G,H) &=&  4 i G^a H^b - 4 i G^{a5} H^{b5} + G^1 H^{ab} + G^{ab} H^1 - 2 G^c H^{d5} \ep_{cd}^{~~ab} \\
&& + 2 G^{c5} H^d \ep_{cd}^{~~ab}- \frac{1}{2}G^5 H^{cd} \ep_{cd}^{~~ab} - \frac{1}{2} G^{cd} H^5
\ep_{cd}^{~~ab} + i G^{ac} H_c^{~b}.
\end{eqnarray}
Using eqs.~(\ref{Amu}) and (\ref{Ftilde}) and the definition of $\ti{F}_{\mu\nu}$, we first derive the components of $\ti{F}_{\mu\nu}$:
\begin{eqnarray}
\ti{F}^1_{\mu\nu} &=& \p_\mu \hat{A}^1_\nu - \p_\nu \hat{A}^1_\mu,\\
\ti{F}^5_{\mu\nu} &=& \p_\mu \hat{A}^5_\nu - \p_\nu \hat{A}^5_\mu + 2 \hat{A}^a_\mu \hat{A}^5_{\nu a} - 2 \hat{A}^a_\nu \hat{A}^5_{\mu a},\\
\ti{F}^a_{\mu\nu} &=& \p_\mu \hat{A}^a_\nu - \p_\nu \hat{A}^a_\mu + \hat{A}^{ab}_\mu \hat{A}_{\nu b} - \hat{A}^{ab}_\nu \hat{A}_{\mu b}
+ 2 \hat{A}^{a5}_\mu \hat{A}^5_\nu - 2 \hat{A}^{a5}_\nu \hat{A}^5_\mu,\\
\ti{F}^{a5}_{\mu\nu} &=& \p_\mu \hat{A}^{a5}_\nu - \p_\nu \hat{A}^{a5}_\mu + \hat{A}^{ab}_\mu \hat{A}^5_{\nu b} - \hat{A}^{ab}_\nu \hat{A}^5_{\mu b}
+ 2 \hat{A}^a_\mu \hat{A}^5_\nu - 2 \hat{A}^a_\nu \hat{A}^5_\mu,\\
\ti{F}^{ab}_{\mu\nu} &=& \p_\mu \hat{A}^{ab}_\nu - \p_\nu
\hat{A}^{ab}_\mu + \hat{A}^{ac}_\mu \hat{A}_{\nu c}^{~~b} -
\hat{A}^{ac}_\nu \hat{A}_{\mu c}^{~~b} + 8 \hat{A}^a_\mu
\hat{A}^b_\nu - 8 \hat{A}^{a5}_\mu \hat{A}^{b5}_\nu,
\end{eqnarray}
and then obtain the components of $\hat{F}_{\mu\nu}$ by using eq.~(\ref{NCF2}):
\begin{eqnarray}
\hat{F}^1_{\mu\nu} &=& \ti{F}_{\mu\nu}^1 + \frac{1}{2} \th^{\a\b} P^1(\{\p_\a \hat{A}_\mu, \p_\b \hat{A}_\nu\})
- \frac{1}{2} \th_{\nu\s} \p_\b \th^{\l\s} \th^{\a\b} P^1(\{\p_\a \hat{A}_\mu, \hat{A}_\l\})\nonumber\\
&& + \frac{1}{2} \th_{\mu\l} \p_\a \th^{\l\s} \th^{\a\b}
P^1(\{\hat{A}_\s, \p_\b \hat{A}_\nu\}) + \frac{1}{2} \th_{\mu\rho} \th_{\nu\s} \p_\d \th^{\rho\a} \p_\e \th^{\s\b} \th^{\d\e} P^1(\{\hat{A}_\a,
\hat{A}_\b\}) \nonumber\\
&&+ \frac{1}{4} \th_{\mu\rho} \th_{\nu\s} \th^{\d\a} \th^{\e\b}
\p_\d \p_\e \th^{\rho\s} P^1(\{\hat{A}_\a, \hat{A}_\b\})+ \frac{i}{2} \th^{\l\d} \p_\l \th_{\mu\rho} \p_\d \th^{\rho\a} \ti{F}^1_{\a\nu}
+ \frac{i}{2} \th^{\l\d} \p_\l \th_{\nu\s} \p_\d \th^{\s\b} \ti{F}^1_{\mu\b} \nonumber\\
&&+ \frac{i}{2} \th^{\l\d} \p_\l \th_{\mu\rho} \th^{\rho\a} \p_\d \ti{F}^1_{\a\nu}
+ \frac{i}{2} \th^{\l\d} \p_\l \th_{\nu\s} \th^{\s\b} \p_\d \ti{F}^1_{\mu\b}
+ \frac{i}{2} \th^{\l\d} \p_\l \th_{\mu\rho} \p_\d \th_{\nu\s} \th^{\rho\a} \th^{\s\b} \ti{F}^1_{\a\b} \nonumber\\
&&+ \th^{\l\d} \p_\l \th_{\mu\rho} \th^{\rho\a} P^1(\hat{A}_\d ,\ti{F}_{\a\nu})
+ \th^{\l\d} \p_\l \th_{\nu\s} \th^{\s\b} P^1(\hat{A}_\d, \ti{F}_{\mu\b}),\label{NCCF1}
\end{eqnarray}
\begin{eqnarray}
\hat{F}^5_{\mu\nu} &=& \ti{F}_{\mu\nu}^5 + \frac{1}{2} \th^{\a\b} P^5(\{\p_\a \hat{A}_\mu, \p_\b \hat{A}_\nu\})
- \frac{1}{2} \th_{\nu\s} \p_\b \th^{\l\s} \th^{\a\b} P^5(\{\p_\a \hat{A}_\mu, \hat{A}_\l\}) \nonumber\\
&&+ \frac{1}{2} \th_{\mu\l} \p_\a \th^{\l\s} \th^{\a\b}
P^5(\{\hat{A}_\s, \p_\b \hat{A}_\nu\})+ \frac{1}{2} \th_{\mu\rho} \th_{\nu\s} \p_\d \th^{\rho\a} \p_\e \th^{\s\b} \th^{\d\e} P^5(\{\hat{A}_\a,
\hat{A}_\b\}) \nonumber\\
&&+ \frac{1}{4} \th_{\mu\rho} \th_{\nu\s} \th^{\d\a} \th^{\e\b} \p_\d \p_\e \th^{\rho\s} P^5(\{\hat{A}_\a, \hat{A}_\b\})
+ \frac{i}{2} \th^{\l\d} \p_\l \th_{\mu\rho} \p_\d \th^{\rho\a} \ti{F}^5_{\a\nu}
+ \frac{i}{2} \th^{\l\d} \p_\l \th_{\nu\s} \p_\d \th^{\s\b} \ti{F}^5_{\mu\b}\nonumber\\
&&+ \frac{i}{2} \th^{\l\d} \p_\l \th_{\mu\rho} \th^{\rho\a} \p_\d \ti{F}^5_{\a\nu}
+ \frac{i}{2} \th^{\l\d} \p_\l \th_{\nu\s} \th^{\s\b} \p_\d \ti{F}^5_{\mu\b}
+ \frac{i}{2} \th^{\l\d} \p_\l \th_{\mu\rho} \p_\d \th_{\nu\s} \th^{\rho\a} \th^{\s\b} \ti{F}^5_{\a\b} \nonumber\\
&&+ \th^{\l\d} \p_\l \th_{\mu\rho} \th^{\rho\a} P^5(\hat{A}_\d ,\ti{F}_{\a\nu})
+ \th^{\l\d} \p_\l \th_{\nu\s} \th^{\s\b} P^5(\hat{A}_\d, \ti{F}_{\mu\b}),\label{NCCF5}
\end{eqnarray}
\begin{eqnarray}
\hat{F}^a_{\mu\nu} &=& \ti{F}_{\mu\nu}^a + \frac{1}{2} \th^{\a\b} P^a(\{\p_\a \hat{A}_\mu, \p_\b \hat{A}_\nu\})
- \frac{1}{2} \th_{\nu\s} \p_\b \th^{\l\s} \th^{\a\b} P^a(\{\p_\a \hat{A}_\mu, \hat{A}_\l\})\nonumber\\
&& + \frac{1}{2} \th_{\mu\l} \p_\a \th^{\l\s} \th^{\a\b} P^a(\{\hat{A}_\s, \p_\b \hat{A}_\nu\})
+ \frac{1}{2} \th_{\mu\rho} \th_{\nu\s} \p_\d \th^{\rho\a} \p_\e \th^{\s\b} \th^{\d\e} P^a(\{\hat{A}_\a, \hat{A}_\b\}) \nonumber\\
&&+ \frac{1}{4} \th_{\mu\rho} \th_{\nu\s} \th^{\d\a} \th^{\e\b} \p_\d \p_\e \th^{\rho\s} P^a(\{\hat{A}_\a, \hat{A}_\b\})
+ \frac{i}{2} \th^{\l\d} \p_\l \th_{\mu\rho} \p_\d \th^{\rho\a} \ti{F}^a_{\a\nu}
+ \frac{i}{2} \th^{\l\d} \p_\l \th_{\nu\s} \p_\d \th^{\s\b} \ti{F}^a_{\mu\b}\nonumber\\
&& + \frac{i}{2} \th^{\l\d} \p_\l \th_{\mu\rho} \th^{\rho\a} \p_\d \ti{F}^a_{\a\nu}
+ \frac{i}{2} \th^{\l\d} \p_\l \th_{\nu\s} \th^{\s\b} \p_\d \ti{F}^a_{\mu\b}
+ \frac{i}{2} \th^{\l\d} \p_\l \th_{\mu\rho} \p_\d \th_{\nu\s} \th^{\rho\a} \th^{\s\b} \ti{F}^a_{\a\b} \nonumber\\
&&+ \th^{\l\d} \p_\l \th_{\mu\rho} \th^{\rho\a} P^a(\hat{A}_\d ,\ti{F}_{\a\nu})
+ \th^{\l\d} \p_\l \th_{\nu\s} \th^{\s\b} P^a(\hat{A}_\d, \ti{F}_{\mu\b}),\label{NCCFa}
\end{eqnarray}
\begin{eqnarray}
\hat{F}^{a5}_{\mu\nu} &=& \ti{F}_{\mu\nu}^{a5} + \frac{1}{2} \th^{\a\b} P^{a5}(\{\p_\a \hat{A}_\mu, \p_\b \hat{A}_\nu\})
- \frac{1}{2} \th_{\nu\s} \p_\b \th^{\l\s} \th^{\a\b} P^{a5}(\{\p_\a \hat{A}_\mu, \hat{A}_\l\}) \nonumber\\
&& + \frac{1}{2} \th_{\mu\l} \p_\a \th^{\l\s} \th^{\a\b}
P^{a5}(\{\hat{A}_\s, \p_\b \hat{A}_\nu\})+ \frac{1}{2} \th_{\mu\rho} \th_{\nu\s} \p_\d \th^{\rho\a}
\p_\e \th^{\s\b} \th^{\d\e} P^{a5}(\{\hat{A}_\a, \hat{A}_\b\}) \nonumber\\
&&+ \frac{1}{4} \th_{\mu\rho} \th_{\nu\s} \th^{\d\a} \th^{\e\b} \p_\d \p_\e \th^{\rho\s} P^{a5}(\{\hat{A}_\a, \hat{A}_\b\})
+ \frac{i}{2} \th^{\l\d} \p_\l \th_{\mu\rho} \p_\d \th^{\rho\a} \ti{F}^{a5}_{\a\nu}
+ \frac{i}{2} \th^{\l\d} \p_\l \th_{\nu\s} \p_\d \th^{\s\b} \ti{F}^{a5}_{\mu\b} \nonumber\\
&&+ \frac{i}{2} \th^{\l\d} \p_\l \th_{\mu\rho} \th^{\rho\a} \p_\d \ti{F}^{a5}_{\a\nu}
+ \frac{i}{2} \th^{\l\d} \p_\l \th_{\nu\s} \th^{\s\b} \p_\d \ti{F}^{a5}_{\mu\b}
+ \frac{i}{2} \th^{\l\d} \p_\l \th_{\mu\rho} \p_\d \th_{\nu\s} \th^{\rho\a} \th^{\s\b} \ti{F}^{a5}_{\a\b} \nonumber\\
&&+ \th^{\l\d} \p_\l \th_{\mu\rho} \th^{\rho\a} P^{a5}(\hat{A}_\d ,\ti{F}_{\a\nu})
+ \th^{\l\d} \p_\l \th_{\nu\s} \th^{\s\b} P^{a5}(\hat{A}_\d, \ti{F}_{\mu\b}),\label{NCCFa5}
\end{eqnarray}
\begin{eqnarray}
\hat{F}^{ab}_{\mu\nu} &=& \ti{F}_{\mu\nu}^{ab} + \frac{1}{2} \th^{\a\b} P^{ab}(\{\p_\a \hat{A}_\mu, \p_\b \hat{A}_\nu\})
- \frac{1}{2} \th_{\nu\s} \p_\b \th^{\l\s} \th^{\a\b} P^{ab}(\{\p_\a \hat{A}_\mu, \hat{A}_\l\})\nonumber\\
&& + \frac{1}{2} \th_{\mu\l} \p_\a \th^{\l\s} \th^{\a\b} P^{ab}(\{\hat{A}_\s, \p_\b \hat{A}_\nu\})
+ \frac{1}{2} \th_{\mu\rho} \th_{\nu\s} \p_\d \th^{\rho\a} \p_\e \th^{\s\b} \th^{\d\e} P^{ab}(\{\hat{A}_\a, \hat{A}_\b\})\nonumber\\
&&+ \frac{1}{4} \th_{\mu\rho} \th_{\nu\s} \th^{\d\a} \th^{\e\b} \p_\d \p_\e \th^{\rho\s} P^{ab}(\{\hat{A}_\a, \hat{A}_\b\})
+ \frac{i}{2} \th^{\l\d} \p_\l \th_{\mu\rho} \p_\d \th^{\rho\a} \ti{F}^{ab}_{\a\nu}
+ \frac{i}{2} \th^{\l\d} \p_\l \th_{\nu\s} \p_\d \th^{\s\b} \ti{F}^{ab}_{\mu\b} \nonumber\\
&&+ \frac{i}{2} \th^{\l\d} \p_\l \th_{\mu\rho} \th^{\rho\a} \p_\d \ti{F}^{ab}_{\a\nu}
+ \frac{i}{2} \th^{\l\d} \p_\l \th_{\nu\s} \th^{\s\b} \p_\d \ti{F}^{ab}_{\mu\b}
+ \frac{i}{2} \th^{\l\d} \p_\l \th_{\mu\rho} \p_\d \th_{\nu\s} \th^{\rho\a} \th^{\s\b} \ti{F}^{ab}_{\a\b}\nonumber\\
&&+ \th^{\l\d} \p_\l \th_{\mu\rho} \th^{\rho\a} P^{ab}(\hat{A}_\d ,\ti{F}_{\a\nu}) + \th^{\l\d} \p_\l \th_{\nu\s} \th^{\s\b} P^{ab}(\hat{A}_\d,
\ti{F}_{\mu\b}), \label{NCCFab}
\end{eqnarray}
where $P(\{G,H\}) \equiv P(G,H) + P(H,G)$.

Substituting all the components of $\hat{F}_{\mu\nu}$ into eq.~(\ref{NCaction}), we finally have the Lagrangian of the noncommutative $U(2,2)$
gauge theory to the first order in $\th^{\mu\nu}$:
\begin{equation}
S = - \int d^4x ~\left(\mathrm{det} \th^{\mu\nu}\right)^{-\frac{1}{2}}~ \ep^{\mu\nu\rho\s} \left(8 \hat{F}^1_{\mu\nu} \hat{F}^5_{\rho\s}
+ \frac{1}{4} \ep_{abcd} \hat{F}^{ab}_{\mu\nu} \hat{F}^{cd}_{\rho\s}\right), \label{NCaction2}
\end{equation}
where $\hat{F}^1_{\mu\nu}$, $\hat{F}^5_{\rho\s}$, and $\hat{F}^{ab}_{\mu\nu}$ are given by eqs.~(\ref{NCCF1}), (\ref{NCCF5}) and (\ref{NCCFab}),
respectively. Although it is not written in an explicit form, this equation is still useful. On the one hand, in light of its formulation
we can discuss the classical limit of the deformed action below. On the other hand, it is the base for us to write down an explicit action
in eq.~(\ref{NCaction3}) after we consider the Seiberg-Witten map.

Strictly speaking, there is no commutative limit of the above action for an arbitrary symplectic tensor because of the
presence of the factor $\left(\mathrm{det} \th^{\mu\nu}\right)^{-\frac{1}{2}}$. But we can deduce the commutative limit for some special
case in which the fluctuation of symplectic tensor is much smaller than the symplectic tensor itself.
We can first take a limit of a constant symplectic tensor and then let the constant tend to zero.
A similar phenomenon also happens~\cite{miao} to the noncommutative $SL(2,C)$ gravity, which might be common in the construction of noncommutative
gravity through the covariant coordinate approach. For more explanations, see the reference.

\subsection{Seiberg-Witten map}
From the above discussions, we can see that there appear additional degrees of freedom besides the vierbein $\hat{e}^a_\mu \equiv \hat{A}^a_\mu$
and the spin connection $\hat{\om}^{ab}_\mu \equiv \hat{A}^{ab}_\mu$. Nevertheless, we can reduce the number of degrees of freedom through imposing
constraints like eq.~(\ref{Constraints})
after we use the so-called ``Seiberg-Witten map"~\cite{SW} which connects noncommutative variables with commutative ones. In principle,
the map can be calculated to any order in $\theta^{\mu\nu}$. However, we
compute it only to the first order for the sake of investigating its primary correction
and of making it consistent with the expansion of star product eq.~(\ref{star}).

For the transformation parameter $\hat{\L}$ and the field $\hat{B}^\mu$, the
map has been provided in ref.~\cite{Calmet:2003} up to the first order in $\th^{\mu\nu}$:
\begin{eqnarray}
\hat{\L} &=& \L + \frac{1}{4} \th^{\mu\nu} \{\p_\mu\L, A_\nu\}, \\
\hat{B^\mu} &=& \th^{\mu\nu} A_\nu - \frac{1}{4}
\th^{\rho\s}\{A_\rho, \p_\s (\th^{\mu\nu} A_\nu) + \th^{\mu\nu}
F_{\s\nu}\}. \label{BMap}
\end{eqnarray}
The map between $\hat{A}_\mu$ and $A_\mu$ can now be obtained from eqs.~(\ref{Borelation}) and (\ref{BMap}):
\begin{equation}
\hat{A}_\mu = A_\mu - \frac{1}{4} \th^{\l\s}\{A_\l, \p_\s A_\mu +
F_{\s\mu}\} - \frac{1}{4} \th_{\mu\nu} \th^{\l\s} \p_\s \th^{\nu\d}
\{A_\l, A_\d\} \label{OmMap},
\end{equation}
which, together with eqs.~(\ref{NCRrelation}), (\ref{NCF1}) and (\ref{NCtheta}), leads to the Seiberg-Witten map of the field strength
also to the first order in $\th^{\mu\nu}$:
\begin{eqnarray}
\hat{F}_{\mu\nu} &= &F_{\mu\nu} + \frac{1}{2} \th^{\a\b}\{F_{\mu\a},
F_{\nu\b}\}
- \frac{1}{4} \th^{\a\b}\{A_\a, (\p_\b + D_\b) F_{\mu\nu}\} - \frac{1}{2} \th_{\nu\s} \p_\a \th^{\rho\s} \th^{\a\b} [F_{\mu\rho}, A_\b] \nonumber\\
&&- \frac{1}{2} \th_{\mu\rho} \p_\a \th^{\rho\s} \th^{\a\b} [F_{\nu\s}, A_\b] + \frac{i}{2} \th^{\a\b} \p_\a \th_{\mu\rho} \p_\b \th^{\rho\s} F_{\s\nu}
+ \frac{i}{2} \th^{\a\b} \p_\a \th_{\nu\s} \p_\b \th^{\rho\s} F_{\rho\mu} \nonumber\\
&& + \frac{i}{2} \th^{\a\b} \th^{\rho\s} \p_\a \th_{\mu\rho} \p_\b
F_{\s\nu}
+ \frac{i}{2} \th^{\a\b} \th^{\rho\s} \p_\a \th_{\nu\s} \p_\b F_{\rho\mu}
+ \frac{i}{2} \th^{\a\b} \p_\a \th_{\mu\rho} \p_\b \th_{\nu\s} \th^{\rho\l} \th^{\s\d} F_{\l\d} \nonumber\\
&&+ \frac{1}{4} \th_{\mu\rho} \th_{\nu\s} \th^{\l\a} \th^{\d\b}
\p_\l \p_\d \th^{\rho\s} \{A_\a, A_\b\}, \label{NCF3}
\end{eqnarray}
where $D_\b F_{\l\s} \equiv \p_\b F_{\l\s} - i [A_\b, F_{\l\s}]$. Now the noncommutative field strength $\hat{F}_{\mu\nu}$
has been expressed by the usual
gauge field and its strength. In order to delete extra degrees of freedom,
we now consider the addition of constraints.\footnote{Theoretically, one can also impose the constraints on
the noncommutative variables, but practically it is hard to solve these constraints and to reduce the number of degrees of freedom.}
As it is convenient to choose similar constraints to that in the commutative case, i.e.
eq.~(\ref{Constraints}), we impose the following constraints in the noncommutative case:
\begin{equation}
A^1_\mu = A^5_\mu = 0, \qquad A^{a5}_\mu = \a A^a_\mu, \qquad F^a_{\mu\nu}= 0.\label{constraint}
\end{equation}
Therefore, with the component formulations of $A_\mu$ and $F_{\mu\nu}$, i.e.
${A}_\mu = {A}^1_\mu + i {A}^5_\mu \g_5 + i {A}^a_\mu\g_a + i {A}^{a5}_\mu \g_a \g_5 + \frac{i}{4} {A}_\mu^{ab}\g_{ab}$ and
$F_{\mu\nu} = F^1_{\mu\nu} + i F^5_{\mu\nu} \g_5 + i F^a_{\mu\nu} \g_a + i F^{a5}_{\mu\nu} \g_a \g_5 + \frac{i}{4} F^{ab}_{\mu\nu} \g_{ab}$,
we simplify $A_\mu$ and $F_{\mu\nu}$ to be
\begin{eqnarray}
A_\mu &=& i A^a_\mu \g_a + i{\a}A^{a}_\mu \g_a \g_5 + \frac{i}{4} A_\mu^{ab} \g_{ab}, \label{67}\\
F_{\mu\nu} &=& \frac{i}{4} F^{ab}_{\mu\nu} \g_{ab} \label{68}.
\end{eqnarray}

Substituting eqs.~(\ref{67}) and (\ref{68}) into $\hat{F}_{\mu\nu}$ (eq.~(\ref{NCF3}))
and decomposing the field strength in terms of the $u(2,2)$ algebra generators, or in an alternative way, i.e.
directly substituting the Seiberg-Witten map of $\hat{A}_\mu$
(eq.~(\ref{OmMap})) into the components of
$\hat{F}_{\mu\nu}$ (eqs.~(\ref{NCCF1})-(\ref{NCCFab})) and then imposing the constraints eq.~(\ref{constraint}) upon the components,
we work out the expansions of the components of
$\hat{F}_{\mu\nu}$ in terms of the usual gauge field and its strength up to the first order in $\theta^{\mu\nu}$:
\begin{eqnarray}
\hat{F}^1_{\mu\nu} &=& \frac{1}{8} \th^{\a\b} F^{ab}_{\mu\a} F_{\nu\b ab} - \frac{1}{2} \th^{\a\b}
\bigg[\left(1 - \a^2\right) A_{\a a} F^{ab}_{\mu\nu} A_{\b b}
+ \frac{1}{4} A_{\a ab} F^{ab}_{\mu\nu} + \frac{1}{8} A_{\a a}^{~~b} F^{ca}_{\mu\nu} A_{\b cb} \nonumber\\
&& - \frac{1}{8} A^a_{\a b} F^{cb}_{\mu\nu} A_{\b ca}\bigg]
+ \frac{1}{2} \th_{\mu\rho} \th_{\nu\s} \th^{\l\a} \th^{\d\b} \p_\l \p_\d \th^{\rho\s} \left[\left(\a^2 -1\right) A^a_\a A_{\b a}
+ \frac{1}{32} A^{ab}_\a A_{\b ab}\right],\label{NCCF12}\\
\hat{F}^5_{\mu\nu} &=& \frac{1}{16} \th^{\a\b} F^{ab}_{\mu\a} F^{cd}_{\nu\b} \ep_{abcd} - \frac{1}{16} \th^{\a\b} \left(A^{ab}_\a \p_\b F^{cd}_{\mu\nu}
+ A^{ab}_\a A^c_{\a e} F^{ed}_{\mu\nu}\right) \ep_{abcd} \nonumber\\
&&+\frac{1}{32} \th_{\mu\rho} \th_{\nu\s} \th^{\l\a} \th^{\d\b} \p_\l \p_\d \th^{\rho\s} A^{ab}_\a A^{cd}_\b \ep_{abcd},\label{NCCF52}\\
\hat{F}^a_{\mu\nu} &=& - \frac{\a}{4} \th^{\a\b} \left(A^d_\a A_{\b e}^{~~c} F^{be}_{\mu\nu} + \frac{1}{2} A^{bc}_\a A_{\b e} F^{de}_{\mu\nu}
- A^d_\a \p_\b F^{bc}_{\mu\nu}\right) \ep_{bcd}^{~~~a} - \frac{i}{2} \th_{\nu\s} \p_\a \th^{\rho\s} \th^{a\b} F^{ab}_{\mu\rho} A_{\b b} \nonumber\\
&&- \frac{i}{2} \th_{\mu\rho} \p_\a \th^{\rho\s} \th^{a\b} F^{ab}_{\nu\s} A_{\b b}
- \frac{\a}{4} \th_{\mu\rho} \th_{\nu\s} \th^{\l\a} \th^{\d\b} \p_\l \p_\d \th^{\rho\s} A^d_\a A^{bc}_\b \ep_{bcd}^{~~~a},\label{NCCFa2}\\
\hat{F}^{a5}_{\mu\nu} &=&  - \frac{1}{4} \th^{\a\b} \left(A^d_\a A_{\b e}^{~~c} F^{be}_{\mu\nu} + \frac{1}{2} A^{bc}_\a A_{\b e} F^{de}_{\mu\nu}
- A^d_\a \p_\b F^{bc}_{\mu\nu}\right) \ep_{bcd}^{~~~a} - \frac{i}{2} \a \th_{\nu\s} \p_\a \th^{\rho\s} \th^{a\b} F^{ab}_{\mu\rho} A_{\b b} \nonumber\\
&&- \frac{i}{2} \a \th_{\mu\rho} \p_\a \th^{\rho\s} \th^{a\b} F^{ab}_{\nu\s} A_{\b b}
- \frac{1}{4} \th_{\mu\rho} \th_{\nu\s} \th^{\l\a} \th^{\d\b} \p_\l \p_\d \th^{\rho\s} A^d_\a A^{bc}_\b \ep_{bcd}^{~~~a},\label{NCCFa52}\\
\hat{F}^{ab}_{\mu\nu} &=& F^{ab}_{\mu\nu} - i \th_{\nu\s} \p_\a \th^{\rho\s} \th^{a\b} F^{ac}_{\mu\rho}A_{\b c}^{~~b}
- i \th_{\mu\rho} \p_\a \th^{\rho\s} \th^{\a\b} F^{ac}_{\nu\s} A_{\b c}^{~~b}
+ \frac{i}{2} \th^{\a\b} \p_\a \th_{\mu\rho} \p_\b \th^{\rho\s} F^{ab}_{\s\nu}\nonumber\\
&&+ \frac{i}{2} \th^{\a\b} \p_\a \th_{\nu\s} \p_\b \th^{\rho\s} F^{ab}_{\rho\mu}
+ \frac{i}{2} \th^{\a\b} \th^{\rho\s} \p_\a \th_{\mu\rho} \p_\b F^{ab}_{\s\nu}
 + \frac{i}{2} \th^{\a\b} \th^{\rho\s} \p_\a \th_{\nu\s} \p_\b F^{ab}_{\rho\mu}\nonumber\\
&&+ \frac{i}{2} \th^{\a\b} \p_\a \th_{\mu\rho} \p_\b \th_{\nu\s} \th^{\rho\l} \th^{\s\d} F^{ab}_{\l\d} \nonumber\\
&\equiv & F^{ab}_{\mu\nu} + F^{ab(1)}_{\mu\nu}\label{NCCFab2},
\end{eqnarray}
where $F^{ab}_{\mu\nu} = R^{ab}_{\mu\nu} + 8 (1 - \a^2) e^a_\mu e^b_\nu$ and $F^{ab(1)}_{\mu\nu}$ denotes the first-order correction to
$\hat{F}^{ab}_{\mu\nu}$. From the component expressions of $\hat{F}_{\mu\nu}$
(eqs.~(\ref{NCCF12})--(\ref{NCCFab2})), we see that the Lagrangian (see eq.~({\ref{NCaction2}})) can be expressed only
by the independent degrees of freedom,
i.e. the vierbein $e^a_\mu \equiv A^a_\mu$ and the spin connection $\om^{ab}_\mu \equiv A^{ab}_\mu$. Moreover, we notice
from eqs.~(\ref{NCCF12}) and (\ref{NCCF52})
that $\hat{F}^1_{\mu\nu}$ and $\hat{F}^5_{\mu\nu}$ contain exclusively the first-order corrections, and as a consequence their product, i.e. the first
term in the Lagrangian (eq.~(\ref{NCaction2})) gives higher up to the second-order contributions in $\th^{\mu\nu}$. Therefore
the first-order corrections to the Lagrangian come only from the second term in
eq.~(\ref{NCaction2}), and thus the action whose Lagrangian is corrected up to the first-order in noncommutative parameters takes the form
\begin{equation}
S = - \frac{1}{4} \int d^4x ~\left(\mathrm{det} \th^{\mu\nu}\right)^{-\frac{1}{2}}~ \ep^{\mu\nu\rho\s} \ep_{abcd} \left(F^{ab}_{\mu\nu} F^{cd}_{\rho\s}
+ F^{ab(1)}_{\mu\nu} F^{cd}_{\rho\s} + F^{ab}_{\mu\nu} F^{cd(1)}_{\rho\s}\right).\label{NCaction3}
\end{equation}
Note that the first-order corrections do not vanish for a general $\th^{\mu\nu}(x)$,
which also occurs~\cite{miao} to the noncommutative $SL(2,C)$ gravity.
If we take the case $\th^{\mu\nu} = {\rm const}$,
which corresponds to the canonical noncommutative spacetime, $F^{ab(1)}_{\mu\nu}$ vanishes.
As a result, the action eq.~(\ref{NCaction3}) reduces to eq.~(\ref{action3}) related to the commutative space just
up to a constant coefficient of proportionality. This coincides with~\cite{Calmet,Mukerjee:2006} the consequence
that the first-order corrections vanish on the canonical noncommutative space.


\section{Conclusion}

In this paper, by following the method of constructing the classical Einstein's gravity from the $U(2,2)$ gauge theory,
we provide a deformed gravity model on a noncommutative space with a  symplectic structure.
In order to obtain the gauge invariant action (eq.~(\ref{NCaction})), we define each quantity by carefully considering its gauge transformation;
see, for instance, eqs.~(\ref{NCBtran}), (\ref{CovCoorTran}), (\ref{NCRtran}) and (\ref{deltatheta}).
Then we calculate the noncommutative field strength to the first order in $\th^{\mu\nu}$
and express it in its components of the $u(2,2)$ algebra generators. Substituting these formulas into the action (eq.~(\ref{NCaction})),
we obtain the noncommutative $U(2,2)$
gauge theory in terms of the noncommutative quantities presented by hats. As the Seiberg-Witten map connects noncommutative quantities
with commutative ones,
we thus use this map to rewrite the noncommutative field strength in terms of commutative quantities
still to the first order in $\th^{\mu\nu}$; see eq.~(\ref{NCF3}).
Furthermore, we impose the constraints (eq.~(\ref{constraint})) and therefore break group $U(2,2)$ to $SO(1,3)$.
Because the noncommutative field strength has been expressed
by the commutative gauge field and its strength, we are able to impose the constraints at the commutative level.
Consequently we obtain the components of the noncommutative field strength in terms of their commutative counterparts; see
eqs.~(\ref{NCCF12})-(\ref{NCCFab2}).
Substituting eqs.~(\ref{NCCF12}), (\ref{NCCF52}) and (\ref{NCCFab2}) into eq.~({\ref{NCaction2}}), we finally give the action of noncommutative gravity.

We note that unlike the commutative theory the first term in eq.~(\ref{NCaction2}) does not vanish in general.
However, this term has no contributions to the Lagrangian when we consider the corrections only up to the first order in $\th^{\mu\nu}$.
As noted in ref.~\cite{miao} for the $SL(2,C)$ gravity, we may not expect the vanishing first-order correction in the $U(2,2)$ case.
Moreover, we do not think the first-order correction in the Lagrangian can be
gauged away. If that was the case, the Riemannian curvature would be
gauged away, too. However, it is impossible for a general curved spacetime.
As a consequence, the result obtained in this paper, though different from that given by ref.~\cite{Banerjee:2007},
coincides with our previous work~\cite{miao} for the $SL(2,C)$ gravity. Furthermore, on the canonical
noncommutative space with constant $\th^{\mu\nu}$, we find that
the first-order correction to the Lagrangian vanishes (see eq.~({\ref{NCaction3}})),
which is consistent with that of refs.~\cite{Calmet,Mukerjee:2006}.


\section*{Acknowledgments}
This work is supported in part by the National Natural Science Foundation of China under Grant No.10675061.

\newpage


\begin{thebibliography}{99}
\bibitem{Snyder}
     H.S. Snyder, {\it Quantized space-time}, Phys. Rev. {\bf 71} (1947) 38; {\it The electromagnetic field in quantized spacetime},
     Phys. Rev. {\bf 72} (1947) 68.

\bibitem{Minwalla}
     S. Minwalla, M. Van Raamsdonk and N. Seiberg, {\it Noncommutative perturbative
     dynamics}, JHEP {\bf 02} (2000) 020 [arxiv:hep-th/9912072].

\bibitem{SW}
   N. Seiberg and E. Witten, {\it String theory and noncommutative geometry}, JHEP {\bf 09} (1999) 032 [arXiv:hep-th/9908142].

\bibitem{Douglas}
    M.R. Douglas and N.A. Nekrasov, {\it Noncommutative field theory}, Rev. Mod. Phys. {\bf 73} (2001) 977
    [arXiv:hep-th/0106048].

\bibitem{Szabo}
    R.J. Szabo, {\it Quantum field theory on noncommutative spaces}, Phys. Rept. {\bf 378} (2003) 207
    [arXiv:hep-th/0109162].

\bibitem{Banerjee}
    R. Banerjee, B. Chakraborty, S. Ghosh,  P. Mukherjee
    and S. Samanta, {\it Topics in noncommutative geometry inspired physics}, Found. Phys. {\bf 39}
    (2009) 1297 [arXiv:0909.1000[hep-th]].

\bibitem{NCgravrev}
    For a recent review, see R.J. Szabo, {\it Symmetry, gravity and noncommutativity}, Class. Quant. Grav. {\bf23} (2006) R199 [arXiv:hep-th/0606233].

\bibitem{Moyal}
    H. Weyl, {\it The theory of groups and quantum mechanics} (Dover, New York, 1931);\\
    H.J. Groenewold, {\it On the principles of elementary quantum mechanics,} Physica {\bf 12} (1946) 405;\\
    J.E. Moyal, {\it Quantum mechanics as a statistical theory,} Proc. Cambridge Phil. Soc. {\bf 45} (1949) 99.

\bibitem{Chamseddine:2000}
    A.H. Chamseddine, {\it Deforming Einstein's gravity}, Phys. Lett. {\bf B 504} (2001) 33 [arXiv:hep-th/0009153].

\bibitem{Marculescu:2008}
    S. Marculescu and F. Ruiz Ruiz, {\it Seiberg-Witten maps for SO(1,3) gauge invariance and deformations of gravity},
    Phys. Rev. {\bf D 79} (2009) 025004 [arXiv:0808.2066[hep-th]].

\bibitem{Chamseddine:2002}
    A.H. Chamseddine, {\it An invariant action for noncommutative gravity in four-dimensions}, J. Math. Phys. {\bf44} (2003) 2534 [arXiv:hep-th/0202137].

\bibitem{Cardellaetal:2002}
    M.A. Cardella and D. Zanon, {\it Noncommutative deformation of four-dimensional Einstein gravity}, Class. Quant. Grav. {\bf20} (2003) L95
    [arXiv:hep-th/0212071].

\bibitem{Chamseddine:2003}
   A.H. Chamseddine, {\it SL(2,C) gravity with complex vierbein and its noncommutative extension}, Phys. Rev. {\bf D 69} (2004) 024015
   [arXiv:hep-th/0309166].

\bibitem{Calmet}
   X. Calmet and A. Kobakhidze, {\it Noncommutative general relativity}, Phys. Rev. {\bf D 72} (2005) 045010
   [arXiv:hep-th/0506157]; {\it Second order noncommutative corrections to gravity}, Phys. Rev. {\bf D 74} (2006) 047702
   [arXiv:hep-th/0605275].

\bibitem{Wess 2005}
   P. Aschieri, C. Blohmann, M. Dimitrijevic, F. Meyer, P. Schupp and J. Wess, {\it A gravity theory on noncommutative spaces},
    Class. Quant. Grav. {\bf22} (2005) 3511 [arXiv:hep-th/0504183].

\bibitem{Wess 2006}
     P. Aschieri, M. Dimitrijevic, F. Meyer and J. Wess, {\it
   Noncommutative geometry and gravity}, Class. Quant. Grav. {\bf23} (2006) 1883 [arXiv:hep-th/0510059].

\bibitem{Banerjee:2007}
    R. Banerjee, P. Mukherjee and S. Samanta, {\it Lie algebraic noncommutative gravity}, Phys. Rev. {\bf D 75} (2007) 125020 [arXiv:hep-th/0703128].

\bibitem{miao}
    Y.-G. Miao and S.-J. Zhang, {\it $SL(2,C)$ gravity on noncommutative space with Poisson
    structure}, Phys. Rev. {\bf D 82} (2010) 084017 [arXiv:1004.2118[hep-th]].

\bibitem{Wessetal:20001}
    J. Madore, S. Schraml, P. Schupp and J. Wess, {\it Gauge theory on
     noncommutative spaces}, Eur. Phys. J. {\bf C 16} (2000) 161
    [arXiv:hep-th/0001203].


\bibitem{Calmet:2003}
X. Calmet and M. Wohlgenannt, {\it Effective field theories on noncommutative spacetime}, Phys. Rev. {\bf D 68} (2003) 025016 [arXiv:hep-ph/0305027].


\bibitem{Kontsevich}
    M. Kontsevich, {\it Deformation quantization of Poisson manifolds, I}, Lett. Math. Phys. {\bf 66} (2003) 157 [arXiv:q-alg/9709040].

\bibitem{Felder:2000}
G. Felder and B. Shoikhet, {\it Deformation quantization with
traces}, Lett. Math. Phys. {\bf 53} (2000) 75 [arXiv:math/0002057];\\
W. Behr and A. Sykora, {\it NC Wilson lines and the inverse Seiberg-Witten map for nondegenerate star products},
Eur. Phys. J. {\bf C 35} (2004) 145 [arXiv:hep-th/0312138];\\
D.V. Vassilevich, {\it Tensor calculus on noncommutative spaces}, Class. Quant. Grav. {\bf 27} (2010) 095020 [arXiv:1001.0766[hep-th]].

\bibitem{Mukerjee:2006}
  P. Mukherjee and A. Saha, {\it A Note on the noncommutative correction to gravity}, Phys. Rev. {\bf D 74} (2006) 027702 [arXiv:hep-th/0605287].


\end{thebibliography}
\end{document}